\documentclass{svproc}
\usepackage{url}
\usepackage{amsmath,amssymb,amsfonts,hyperref,mathtools}
\usepackage{subfigure}
\usepackage{graphicx}
\usepackage{orcidlink}
\usepackage{algorithmic,algorithm2e}

\begin{document}
\mainmatter              
\title{Active Restoration of Lost Audio Signals using Machine Learning and Latent Information}

\titlerunning{ARLAS}  
%
\author{Zohra Adila Cheddad\inst{1} \and Abbas Cheddad \orcidlink{0000-0002-4390-411X} \inst{2}}
\authorrunning{Z.A. Cheddad and A. Cheddad} 
%
\tocauthor{.}
\institute{Département of Mathematics, Université Frères Mentouri I, 250 17 Constantine, Algeria.\\
\email{zcheddad929@gmail.com}
\and
Department of Computer Science, Blekinge Institute of Technology, 371 79 Karlskrona, Sweden.\\ 
\email{abbas.cheddad@bth.se}\\
\texttt{https://abbascheddad.net}}
\maketitle              

\begin{abstract}
Digital audio signal reconstruction of a lost or corrupt segment using deep learning algorithms has been explored intensively in recent years. Nevertheless, prior traditional methods with linear interpolation, phase coding and tone insertion techniques are still in vogue. However, we found no research work on reconstructing audio signals with the fusion of dithering, steganography, and machine learning regressors. Therefore, this paper proposes the combination of steganography, halftoning (dithering), and state-of-the-art shallow and deep learning methods. The results (including comparing the SPAIN, Autoregressive, deep learning-based, graph-based, and other methods) are evaluated with three different metrics. The observations from the results show that the proposed solution is effective and can enhance the reconstruction of audio signals performed by the side information (e.g., Latent representation) steganography provides.
Moreover, this paper proposes a novel framework for reconstruction from heavily compressed embedded audio data using halftoning (i.e., dithering) and machine learning, which we termed the HCR (halftone-based compression and reconstruction). This work may trigger interest in optimising this approach and/or transferring it to different domains (i.e., image reconstruction). Compared to existing methods, we show improvement in the inpainting performance in terms of signal-to-noise ratio (SNR), the objective difference grade (ODG) and Hansen's audio quality metric. In particular, our proposed framework outperformed the learning-based methods (D2WGAN and SG) and the traditional statistical algorithms (e.g., SPAIN, TDC, WCP). \footnote{CITE AS:
Cheddad, Z.A., Cheddad, A. (2024). Active Restoration of Lost Audio Signals Using Machine Learning and Latent Information. In: Arai, K. (eds) Intelligent Systems and Applications. IntelliSys 2023. Lecture Notes in Networks and Systems, vol 822. Springer, Cham. https://doi.org/10.1007/978-3-031-47721-8\_1}

\end{abstract}
\section{Introduction}
Corrupt audio files and lost audio transmission and signals are severe issues in several audio processing activities such as audio enhancement and restoration. For example, in different applications and music enhancement and restoration situations, gaps could occur for several seconds~\cite{Ebner20}. Audio signal reconstruction remains a fundamental challenge in machine learning and deep learning despite the remarkable recent development in neural networks~\cite{Ebner20}. Audio in-painting, audio interpolation/extrapolation, or waveform replacement address the restoration of lost information in audio. The reconstruction aims to provide consistent and relevant information while eliminating audible artefacts to keep the listener unaware of any occurring issues~\cite{Marafioti19}.
The active reconstruction can be considered a preemptive security measure to allow for self-healing when part of the audio becomes corrupted. Active reconstruction means reconstructing lost signals by incorporating side information retrieved from pre-embedded data. Thus, the technique is not intended for any degraded audio but only for audio protected by the steganography information. To this end, and to the best of our knowledge, we found no prior research work on the active reconstruction of audio signals with the fusion of steganography (an information hiding technique), halftoning and machine learning (ML) models. The initial idea (without ML) was proposed in a PhD thesis as an application of steganography. The hiding strategy of steganography can be tailored to act as an intelligent streaming audio/video system that uses techniques to conceal transmission faults from the listener that are due to lost or delayed packets on wireless networks with bursty arrivals; thus, providing a disruption tolerant broadcasting channel ~\cite{Cheddad2009}.

While it is possible to apply passive recovery procedures to estimate missing audio segments, these procedures, whether they are based on statistical inferences or deep learning, would be futile when the missing gap exceeds half a second, thus, prohibiting their use in a real-world scenario. Hence, the corollary to that is our recommendation in this work to harness the power of machine learning methods by embedding compressed latent information.
In addition to its ultimate purpose of audio recovery, our proposed approach also benefits security systems in protecting audio files from unauthorised manipulation. It may also be extended to image inpainting.

The contribution of this work is fourfold: 

\textemdash{A halftone-based compression and reconstruction (HCR).}

\textemdash{Orchestration of the three scientific disciplines: steganography, compression, and audio processing, for audio reconstruction or inpainting.}

\textemdash{A new framework that enables sequence-to-sequence deep learning (Seq2Seq) and Random Forest (RF), to locally train on other intact segments and latent representation.}

\textemdash{Unlike existing methods (traditional or deep-learning based), our approach can handle extensive gap reconstruction (e.g., 4000ms, 8000ms), more information in Sec \ref{sec:Results}.}

The remainder of the paper is divided as follows: Sec. \ref{RW} discusses related work. Sec. \ref{sec:Method} furnishes the necessary details regarding our proposed method, which we termed ARLAS, including our approach of imposing heavy compression on audio signals. The experimental set-up is highlighted in Sec. \ref{sec:ES}. Sec. \ref{sec:Results} presents the results and compares the performance against existing competitive methods. This paper concludes with some notes in Sec. \ref{sec:Conc}.

\section{Related Work} \label{RW}
In the work of Khan et al.~\cite{Khan20}, the authors proposed a modern neuro-evolution algorithm, Enhanced Cartesian Genetic Programming Evolved Artificial Neural Network (ECGPANN), as a predictor of the lost signal samples in real-time. The authors have trained and tested the algorithms on audio speech signal data and evaluated them on the music signal. A deep neural network (DNN)-based regression method was proposed in~\cite{Lee15} for a packet loss concealment (PLC) algorithm to predict a missing frame's characteristics. Two other DNNs were developed for the training by integrating the log-power spectra and phases based on the unsupervised pre-training and supervised fine-tuning. The algorithm then provides the previous frame features to the trained DNN to reconstruct the missing frames. In~\cite{Ebner20}, researchers have analyzed audio gaps (500 - 550 ms) and used Wasserstein Generative Adversarial Network (WGAN) and Dual Discriminator WGAN (D2WGAN) models to reconstruct the lost audio content. In Khan et al.~\cite{Khan17}, the authors proposed an audio signal reconstruction model called Cartesian Genetic Programming evolved Artificial Neural Network (CGPANN), which was more efficient than the interpolation-extrapolation techniques. The developed model was robust in recovering signals contaminated with up to 50\% noise. In~\cite{Marafioti19}, the authors proposed a DNN structure to restore the missing audio content based on the audio gaps. The signals provided in the audio gaps in the DNN structure were time-frequency coefficients (either complex values or magnitude). In the work of Sperschneider et al.~\cite{Sperschneider15}, the authors presented a delay-less packet-loss concealment (PLC) method for stationary tonal signals, which addresses audio codecs that utilizes a modified discrete cosine transformation (MDCT). In the case of a frame loss, tonal components are identified using the last two obtained spectra and their pitch information. Furthermore, the MDCT coefficients of the tonal components were estimated using phase prediction based on the detection of tonal components. Mokrý et al.~\cite{Mokry19} presented an inpainting algorithm called SPAIN (SParse Audio INpainter) developed by an adaptation of the successful declipping method, SPADE~\cite{Kitic15}, to the context of audio inpainting. The authors show that the analysis of their algorithm, SPAIN, performs the best in terms of SNR (signal-to-noise ratio) among sparsity-based methods and reaches results on a par with the state-of-the-art Janssen algorithm~\cite{Janssen86} \footnote{Iteratively fits autoregressive models using a gap's all previous points for forward estimation and all its future points for backward estimation.} for audio inpainting. A composite model for audio enhancement that combines the Long Short-Term Memory (LSTM) and the Convolutional Neural Network (CNN) models was proposed in~\cite{Hasannezhad21}. Perraudin et al. ~\cite{Perraudin18}, proposed a reconstruction method for the loss of long signals in audio (i.e., Music signals). The concealment of such signal defects is based on a graph that encodes signal structure in terms of time-persistent spectral similarity and an intuitive optimization embedding scheme. Mokrý and Rajmic ~\cite{Mokry20} proposed a heuristic method for compensating for energy loss after running the $ \mathcal{L}_1 $ minimization. Their idea is to take the solution and modify it by entrywise multiplication of the recovered gap in the time domain by a compensation curve in order to increase its amplitude; they termed this approach the Time Domain Compensation (TDC) algorithm.
\section{ARLAS: Active Restoration of Lost Audio Signals}
\label{sec:Method}
Our method aims to reconstruct a realistic segment from audio containing corrupted or dropped regions using active embedding and machine learning. In this section, we describe the proposed method and discuss the individual stages in more detail.
\subsection{Halftone-based Compression and Reconstruction (HCR)}
In the field of bit-rate reduction, or data compression, the Lempel–Ziv (LZ) or its variant, the Lempel–Ziv–Welch (LZW), are popular methods. However, they are slow and prone to failure if data corruption occurs, as in our case, there were more than 700 samples deleted from the audio signal. Therefore, finding a compression method that is more immune to data corruption and can provide heavy compression and good approximate reconstruction is desired. Hence, the proposed HCR is meant to exploit the halftoning for this purpose. The algorithm this work adopts is that of \textit{Floyd Steinberg}, which applies error forward diffusion~\cite{Floyd76}. The rationale behind conceiving the notion of HCR is that the embedding of data in the least significant bits (LSB) of a bit-stream, necessitates dealing with binary data. 
Let the original audio sampled data be denoted by the vector $\overrightarrow{v}$ ($\forall v_i \in\mathbb{R}$), which is then transformed into a matrix $\textbf{D}$ where $v_{xy}\in \textbf{D} _ {n,m}$ (the $x$-th row and $y$-th column in $\textbf{D}$). The matrix $\textbf{D}$ is filled column-wise with suitable dimensions whose automatic estimation is outside of the scope of this work, see equation~\ref{eq1}.

\begin{equation} \label{eq1}
\begin{multlined}
\overrightarrow{v} = ([v_1, v_2, v_3,...,v_{j-1}, v_j])^T \mapsto \textbf{D}
= \\
\begin{bmatrix} 
    v_{11} & v_{12} &\dots &  v_{1m} \\
    v_{21} & v_{22} &\dots &  v_{2m} \\
    \vdots & \vdots & \ddots & \vdots\\
    v_{n1} &  v_{n2} &  \dots   & v_{nm} \\
    \end{bmatrix}
\end{multlined}
\end{equation}

\begin{figure}
\centering
\subfigure[]{\includegraphics[width=28mm]{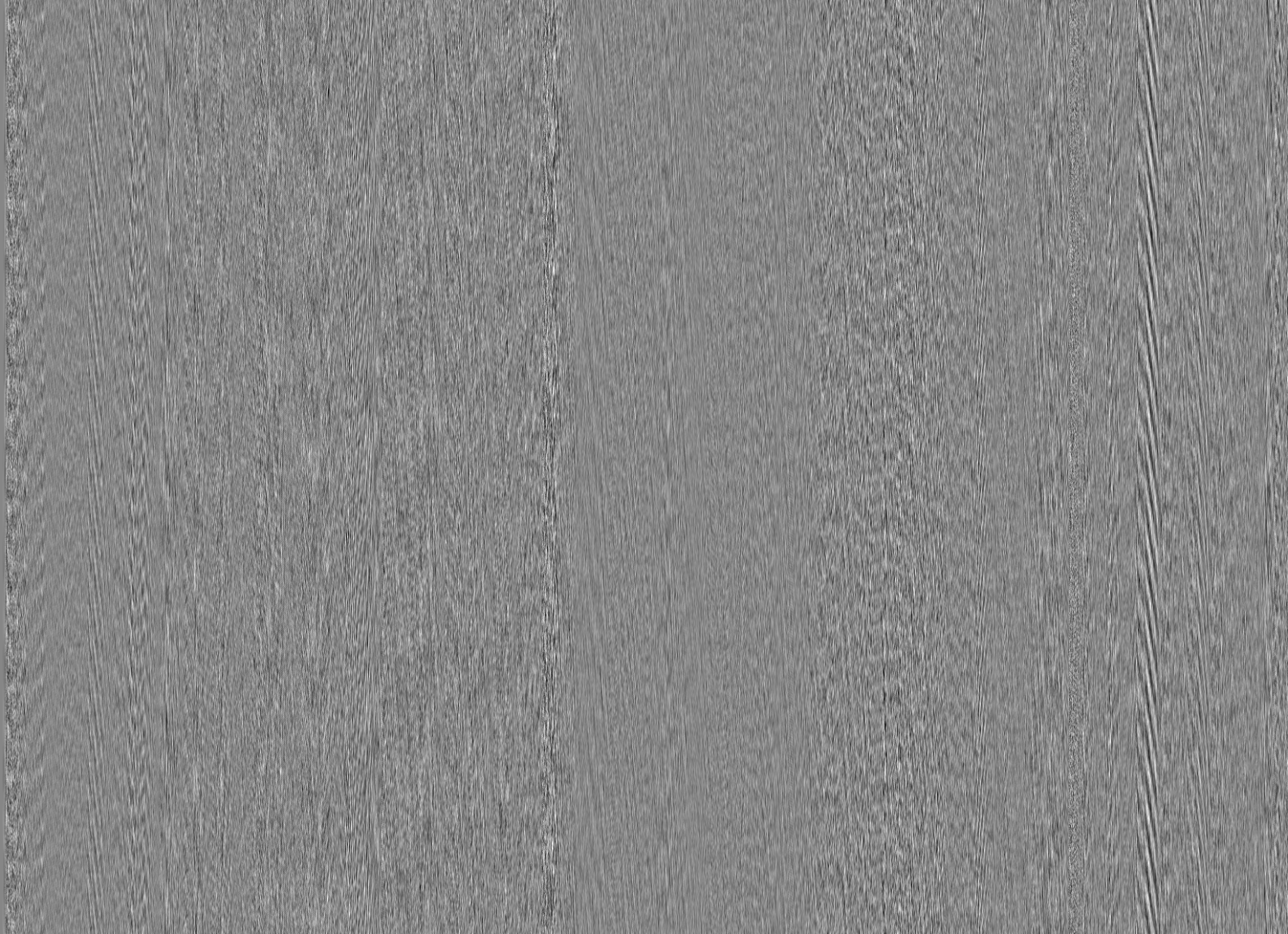}}\hspace{0.01 mm}
\subfigure[]{\includegraphics[width=28mm]{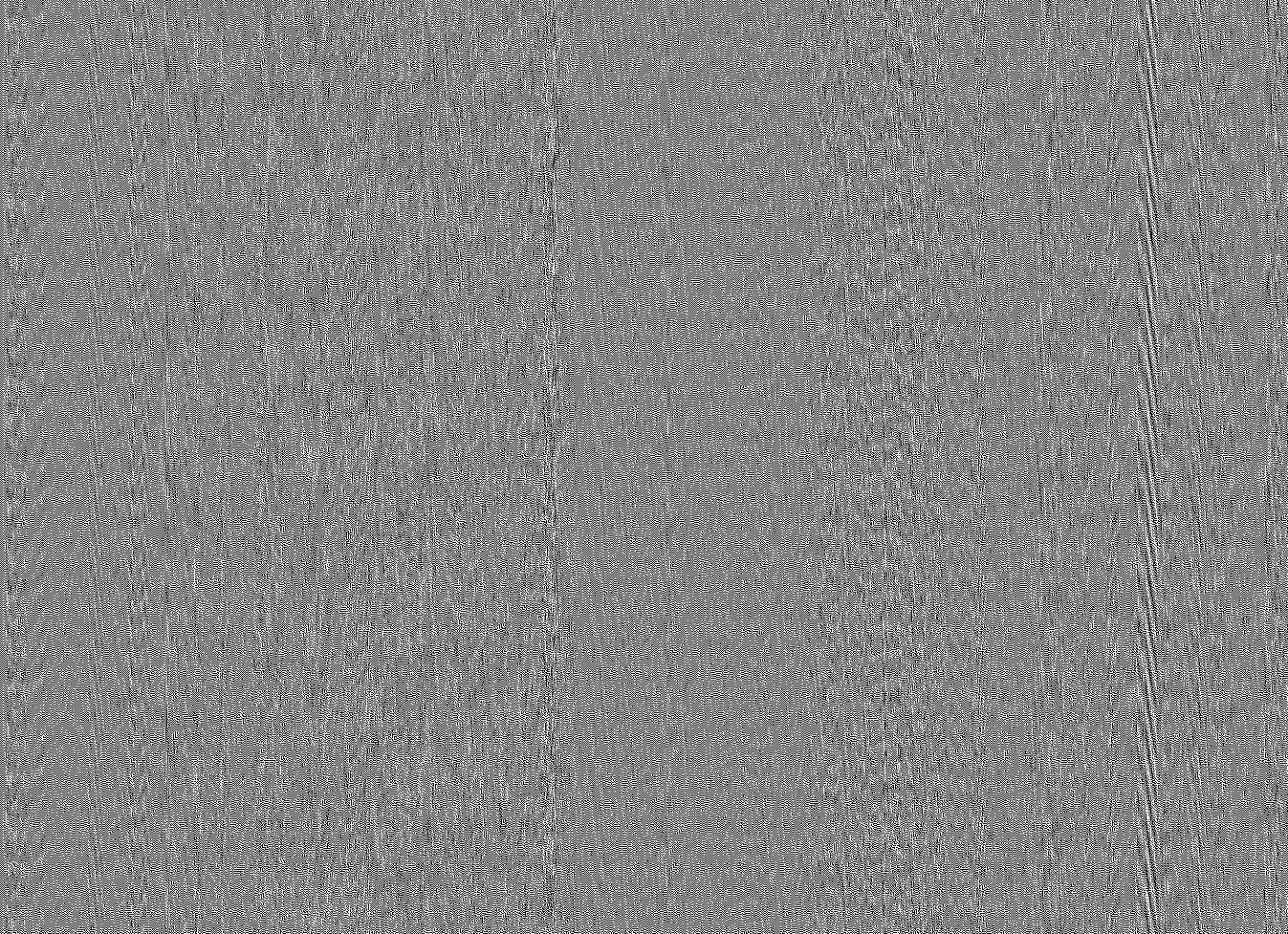}}\hspace{0.01 mm}
\subfigure[]{\includegraphics[width=28mm]{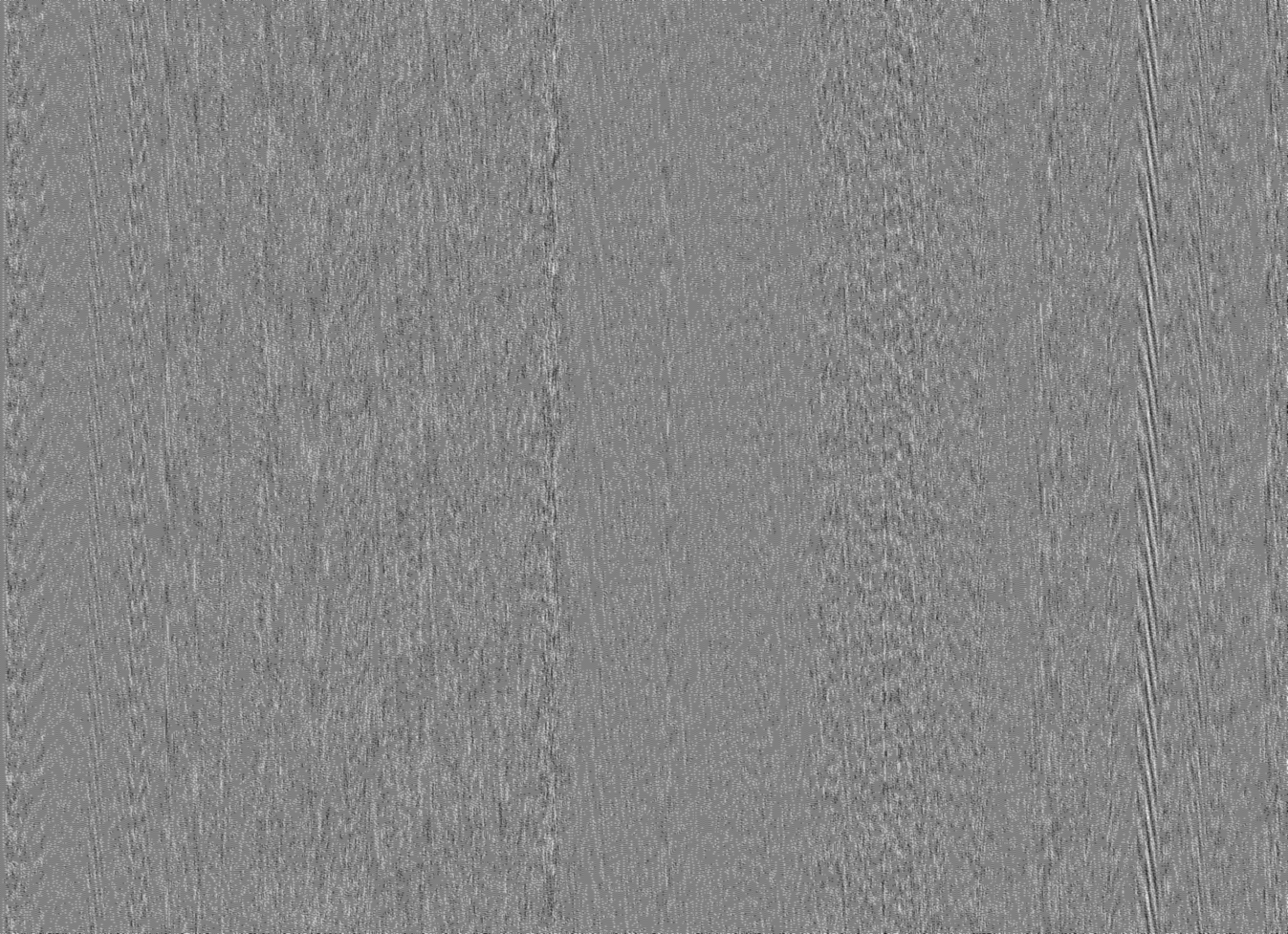}}
\\
\subfigure[]{\includegraphics[width=85mm]{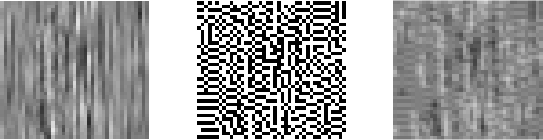}}
\caption{HCR Visual Inspection: (a) Original audio data reshaped using Eq.~\ref{eq1} and visualised (b) Halftone of \textit{(a)} (binary image), (c) Reconstructed \textit{(a)} from \textit{(b)} and (d) Small patches cropped from each image left to right, respectively.} 
\label{fig1}
\end{figure}

\begin{figure}
\centering
\includegraphics[width=85mm]{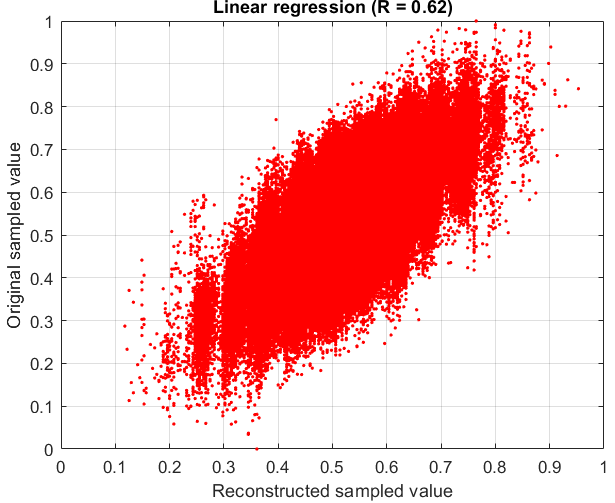}
\caption{A dot plot depicting the correlation between the original data and its estimated version. Note that the reconstruction is made from merely binary values (without ML and signal drop). This figure shows how much of the information is lost in the compression process; however, we rely on our deep learning model to learn reconstructing the original signal from this approximation.} 
\label{fig2}
\end{figure}

The matrix $\textbf{D}$ (remapped to 8bit unsigned integers) is then passed to the dithering phase using Floyd Steinberg algorithm, which results in a binary matrix (as seen in Fig.~\ref{fig1}b) that could be partially inverted. This contributes to the heavy compression that we obtain. For instance, the original audio sampled data and the resulting compressed vector pertaining to Fig.~\ref{fig1}b (vectorized), show the following properties: Original Audio (1316019  | \textbf{10.04 MB}  | Byte) and its corresponding compressed vector (1316019   |  \textbf{1.26 MB}   |   Binary), for the length, size and type, respectively.
The error diffusion algorithm exploits the optical system of the human eye which acts as a low-pass filter removing all high frequencies resulting in the illusion of perceiving a dithered image (only binary) as a continuous tone image. Hence, it follows from such notation that in order to partially inverse the dithering operation we need to apply a low-pass filter to attenuate high frequencies (in our case, we choose a 2D Gaussian filter as in Eq.~\ref{eq2} with a kernal size $ = 2\times \lfloor 2\times\sigma \rfloor+1 $).
\begin{equation} \label{eq2}
G(x,y)={\frac {1}{2\pi \sigma ^{2}}}e^{-{\frac {x^{2}+y^{2}}{2\sigma ^{2}}}}
\end{equation}
where $ \sigma = 1.50 \ $ (determined empirically) and $x$, $y$ are the coordinates of the matrix $\textbf{D}$ where $x \in (1,..,n)$ and $y \in (1,..,m)$. Recently, the development of deep machine learning rekindled interest in addressing the inverse halftoning  problem by optimization-based filtering ~\cite{Kim18}~\cite{LiDeep16}. Nevertheless, we opt to use the simple and computationally-inexpensive method insinuated above.
In order to scrutinize the efficiency of the reconstruction, we calculated the correlation between the original sampled data against the estimated values from the above HCR process. The reconstruction still demonstrates a good correlation R = 0.62, despite the dithered version constitutes only binary values (either 0 or 1); see Fig.~\ref{fig2}. The fitted linear regression model is depicted in Table~\ref{tab1}. In Fig.~\ref{fig1}c, we observe that the process captures a noisy structure and orientation of the original data shown in Fig.~\ref{fig1}a; therefore, when this side-information is wedded to machine learning (particularly the state-of-the-art models), it leverages the quality of audio the algorithm reconstructs. This was the impetus for the initiation of this study. To gauge this performance improvement, we tested some ML models whose results appear in section~\ref{ML}.
\begin{table}
\caption{Effect-estimates and \textit{P-values} (Wald tests) from fitting a linear regression model for the data plotted in Fig.~\ref{fig1}. Number of observations: 1316019, Error degrees of freedom: 1316017 Root Mean Squared Error: 0.0527 R-squared: 0.384,  Adjusted R-Squared: 0.384.}\label{tab1}
\begin{center}
\begin{tabular}{|l|l|l|l|l|}
\hline
& Estimate &  SE & tStat & P-value\\
\hline
{\bfseries (Intercept)} &  0.23791 & 0.00032051 & 742.29 & 0\\
{\bfseries x1} &  0.54719  &   0.00060376  &   906.3  &    0\\
\hline
\end{tabular}
\end{center}
\end{table}
\subsection{Deployed deep/machine learning architectures} \label{ML}
In this section, we list down the different machine learning models we deploy in this study. 
\subsubsection{Shallow machine learning}
\textit{Random Forest (RF)} - RF regressor is a supervised decision learning technique for regression that employs the ensemble learning method. The most important parameter is the number of trees which is set to 100 as per the recommendation in the literature~\cite{Dasari19}. It has been demonstrated that RF is more robust than other similar approaches for handling small samples, high dimensional and nonlinear fitting~\cite{Dasari2020}~\cite{Espinosa21}.
\subsubsection{Deep machine learning}
\textit{Long short-term memory (LSTM)} - LSTM is a type of recurrent neural network (RNN) models; it is best suited to make predictions based on time series data by learning long-term dependencies~\cite{SunDu17}. We have implemented a multi-layer LSTM recurrent neural network to predict the missing signal values with the Keras TensorFlow library (more details in Sec.~\ref{sec:SoftE}).
The LSTM is inherently able to learn some long-term dependencies and find patterns over time, which ultimately makes its next prediction more accurate. The latter distinguishes it from traditional shallow ML models. The LSTM also overcomes the vanishing gradient problem that other RNN-based architectures are prone to ~\cite{Graves12}. This is the impetus for using LSTM in audio processing, whose signals can be viewed as time-series inputs.
\subsubsection{Training and testing segments}
\begin{figure*}
\centering
\includegraphics[width=\textwidth]{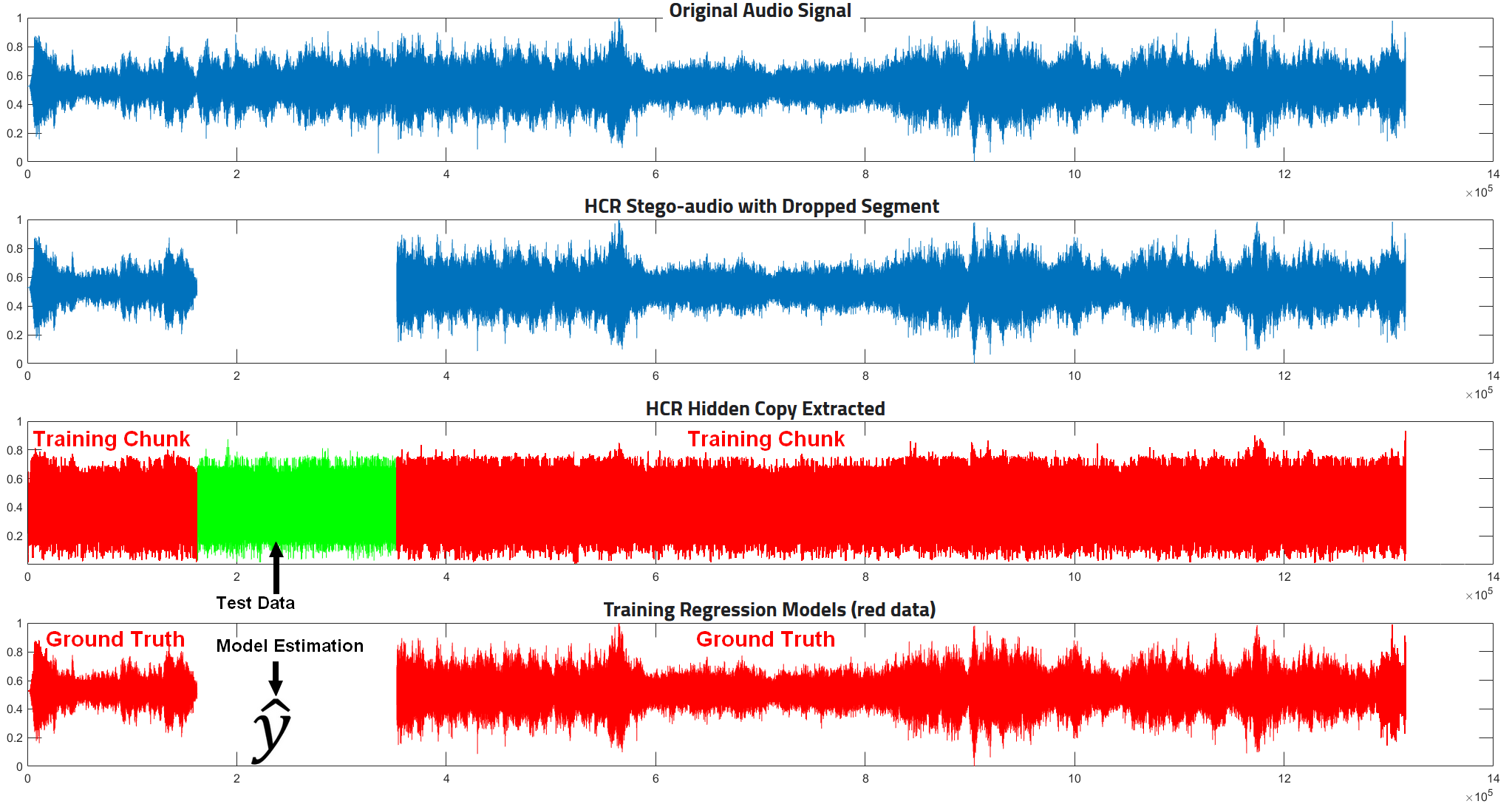}
\caption{Audio track segments to train (red) and test (green) machine learning models.} \label{fig3}
\end{figure*}

This section briefly discusses how the two deep/machine learning models are trained and tested.
In Fig.~\ref{fig3} (top row), the original audio sampled signal is displayed for mere comparison. Fig.~\ref{fig3} (second row) shows the stego-audio signal with the embedded copy and a simulation of signal loss (i.e., empty segment). Note that the stego-audio \footnotetext{In the domain of Steganography, a stego-carrier is a signal similar to the original one except that it carries in it concealed data.} and the original audio look identical since all what was flipped is the last LSB value. Subsequently, we extract the hidden data from the LSB plane, and the values in the extracted binary vector are rearranged by using the same secret key (e.g., for simplicity, we choose the length of the audio segment as the key). The vector is then transformed to a matrix using Eq. \eqref{eq1} (which should correspond to the dithered version), and then a 2D Gaussian filter is applied using Eq. \eqref{eq2}. The result is then vectorised to yield the plot shown in Fig.~\ref{fig3} (third row). Part of this vector will be used for training and validation (i.e., the red segments), and the tuned model is finally applied to the test data (i.e., the green segment) to predict (reconstruct) the lost segment.
 Performing global tuning based on locally adaptive learnt statistics was discussed in the literature though in a different context ~\cite{Yogarajah12}.
 Fig.~\ref{fig4}, along with the Algorithms (~\ref{alg:One}, ~\ref{alg:Two}, ~\ref{alg:Three} ), provide an in-depth detailed description of our approach. In essence, Alg.~\ref{alg:One} describes our embedding approach, while  Alg.~\ref{alg:Two} provides feature extraction based primarily on the extracted embedded data and on its derived augmented feature space using scalogram (namely, the continuous 1-D wavelet transform)~\cite{Lilly17}, which shall form the backbone for the training of the machine learning models.
 \SetKwComment{Comment}{/* }{ */}
\RestyleAlgo{ruled}
\begin{algorithm}
\caption{HCR-Embedding Algorithm}\label{alg:One}
\KwData{$Audio \hspace{0.2cm} file \hspace{0.2cm} S$}
\KwResult{$Self-embedded \hspace{0.2cm} file \hspace{0.2cm} SS$}
\setlength\parindent{5pt}
$S^\prime \gets $ Scale the vector S to the interval [[0,1]*255]\;
\setlength\parindent{15pt} $S^{\prime\prime} \gets $ Reshape $S^\prime$ into a matrix\ \Comment*[r]{dimensions are calculated adaptively}
  $D \gets $ Perform image dithering on $S^{\prime\prime}$\;\Comment*[r]{heavy lossy compression}
$D^\prime \gets $ Flatten D into a vector\;
SET SEED $ \alpha = length(D^\prime)$\;
$D^{\prime\prime} \gets $ Permute bit stream positions of $D^\prime$ with $\alpha$\; \Comment*[r]{to avoid localised distortion}
\setlength\parindent{5pt}
$SS \gets $ embed $D^{\prime\prime}$  into LSB (least significant bits) of $S^\prime$\;
$SS \gets $ rescale(SS)\;
\setlength\parindent{0pt}
$ \textbf{Return} \hspace{0.2cm} SS$\;
\end{algorithm}
\SetKwComment{Comment}{/* }{ */}
\RestyleAlgo{ruled}
\begin{algorithm}
\caption{HCR-Feature Extraction Algorithm}\label{alg:Two}
\KwData{$Self-embedded \hspace{0.2cm} file \hspace{0.2cm} SS \hspace{0.2cm}with\hspace{0.2cm} gap\hspace{0.2cm} G$}
\KwResult{$Feature \hspace{0.2cm}descriptors \hspace{0.2cm} file \hspace{0.2cm} F$}
\setlength\parindent{5pt}
$S^\prime \gets $ Scale the vector SS to the interval [[0,1]*255]\;
$S^{\prime\prime} \gets $ extract the LSB bit stream of $S^\prime$\;
SET SEED $ \alpha = length(S^{\prime\prime})$\;
$E \gets $ Invert permutation of the bit stream positions of $S^{\prime\prime}$ with $\alpha$\;
$E^\prime \gets$ Reshape $E$ into a matrix \;
$E^{\prime\prime} \gets $ Apply 2D Gaussian filter as in Eq.~\ref{eq2} and flatten it\;
$W \gets $ Compute scalogram from $E^{\prime\prime}$  using continuous 1-D wavelet transform and $ \mathcal{L}_1 $ optimisation\;
$F \gets \hspace{0.2cm} concat_{vertically} \hspace{0.2cm} \Re(W) \& E^{\prime\prime} $\;
\Comment*[r]{to form the augmented feature space}
$ \textbf{Return} \hspace{0.2cm} F$\;
\end{algorithm}
\SetKwComment{Comment}{/* }{ */}
\RestyleAlgo{ruled}
\begin{algorithm}
\caption{HCR-RF/LSTM Algorithm}\label{alg:Three}
\KwData{$Self-embedded \hspace{0.2cm} file \hspace{0.2cm} SS \hspace{0.2cm}with\hspace{0.2cm} gap\hspace{0.2cm} G $}
\KwData{$Feature \hspace{0.2cm}descriptors \hspace{0.2cm} file\hspace{0.2cm} F$}
\KwResult{$Reconstructed \hspace{0.2cm} audio \hspace{0.2cm} file \hspace{0.2cm} \hat{S}$}
$M \gets length(SS)$\;
$N \gets 1$\;
$TrainX \gets [\hspace{0.1cm}]$\; 
$TrainY \gets [\hspace{0.1cm}]$\;
$TestX \gets [\hspace{0.1cm}]$\;
$TestY \gets [\hspace{0.1cm}]$\;
\While{$N \leq M$}{
  \eIf{$N$ $\in G$}{
    $TestX \gets F[N]$ \Comment*[r]{Extract features pertaining to the gap G}
  }{\If{$N$ $\notin G$}{
      $TrainX \gets F[N]$\;
      $TrainY \gets SS[N]$ \Comment*[r]{Response variables are those regions of the uncorrupt signal SS}
    }
  }
}
$HCRModel \gets $ Train  RF/LSTM using \{TrainX, TrainY\}\;
$\hat{G} \gets $  $HCRModel$ \{TestY\}\Comment*[r]{Estimate the gap using the trained model}
$\hat{S} \gets SS + \hat{G}$\;
$ \textbf{Return} \hspace{0.2cm} \hat{S}$
\end{algorithm}
\section{Experimental Set-up}
\label{sec:ES}
The experimental samples in this paper are commonly used audio file samples exhibiting different music instruments. As this research focuses on active reconstruction, the audio input is in \textit{.wav} file format (PCM 16 bit). The main reason for selecting this format is the sound quality, as it preserves the originality of analogue audio without any compression.
\subsection{State of the Art methods}
The state-of-the-art sparsity-based inpainting techniques to which we compared our approach are the non-learning methods {'ASPAIN'~\cite{Mokry19},	Chambolle–Pock (CP)~\cite{Chambolle11}, 	Douglas–Rachford based (DR)~\cite{Combettes11},	Janssen Autoregressive~\cite{Janssen86}, Orthogonal Mathching Pursuit (OMP)~\cite{Adler12},	'SSPAIN-H'~\cite{Mokry19},	Revisited and re-weighted methods ('reCP', 'reDR', 'wCP', 'wDR')~\cite{Mokry20}, and the Time Domain Compensation (TDC)~\cite{Mokry20}}. Moreover, we also included a deep-learning based method (D2WGAN)~\cite{Ebner20} inspired from ~\cite{Donahue19}, and a graph-based (Similarity Graphs -SG-)~\cite{Perraudin18}. Our approaches are labeled 'RF', and 'LSTM'. The RF and the LSTM are trained on the extracted pre-embedded data to perform the reconstruction of the missing gap.
\subsection{Data set (audio samples)}  \label{sec:Dataset}
We used the same audio benchmark as in ~\cite{Mokry19}, which comprises 10 files:\textit{ Violin, Clarinet, Bassoon, Harp, Glockenspie, Celesta, Accordion, Guitar sarasate, Piano schubert, and Wind ensemble stravinsky}. The gap length of 100 ms and 300 ms were chosen randomly from the set so that the gaps do not overlap. Therefore, we ran a total of 280 tests $(14 (methods)\times10(segments)\times2(gaps))$. These tests (Phase I) were run on all methods reported in the previous section except the D2WGAN and SG.
In the second stage (Phase II), we deploy the deep learning and the SG methods. We first select the top-performing methods from Phase I. Then, we perform additional tests using the ~\textit{Piano} and the ~\textit{Acoustic Guitar and Orchestra} (Mixed instruments) audio files, the gaps' length of 520 ms, 820 ms, 865ms, and 883ms were chosen following the paper ~\cite{Ebner20}.
\subsection{Evaluation metrics} \label{Metrics}
We utilise three commonly used reference-based metrics to evaluate the performance of the different audio restoration algorithms. The first is the objective difference grade (ODG) which is described in the Perceptual Evaluation of Audio Quality (PEAQ) standard algorithm used for objective measurements of perceived audio quality. It is believed to be based on generally accepted psycho-acoustic principles that approximate the subjective difference grade used in human-based audio tests ~\cite{Thiede00}~\cite{Kabal02}~\cite{Huber06}. The second is related to ODG but uses Hansen's method and a different model for speech quality estimation~\cite{Hansen00}; this metric is termed herein    "\textit{QC}". The third performance measure is the scale-invariant source-to-noise ratio (SI-SNR), which has recently been advocated for as the preferred evaluation metric in lieu of the standard source-to-distortion ratio (SDR), SI-SNR (hereforth abbreviated as SNR) is defined as follows~\cite{Luo19}:
\begin{equation}\label{eq3}
{\begin{cases} {\mathbf{s}}_{target} := \frac{\langle \hat{ {\mathbf{s}}}, {\mathbf{s}} \rangle {\mathbf{s}}}{\left\Vert {\mathbf{s}} \right\Vert ^2}\\
{\mathbf{e}}_{noise} := \hat{ {\mathbf{s}}} - {\mathbf{s}}_{target}\\
\text{SI-SNR} := 10\,log_{10}\frac{\left\Vert {\mathbf{s}}_{target} \right\Vert ^2}{\left\Vert {\mathbf{e}}_{noise} \right\Vert ^2} \end{cases}}
\end{equation}
where $\hat{{\mathbf{s}}} \in \mathbb{R}^{1\times T}$ and ${\mathbf{s}} \in \mathbb {R}^{1\times T}$ are the reconstructed and original clean sources, respectively, and $ \left\Vert {\mathbf{s}} \right\Vert ^2 = \langle {\mathbf{s}}, {\mathbf{s}} \rangle $  denotes the signal power. Scale invariance is ensured by normalizing to zero-mean prior to the calculation.
All of the metrics capture objective measures of signal quality. The ODG metric's values range from 0.0 (imperceptible audio distortion) to -4.0 (very annoying distortion). The SNR is measured in dB (decibel); the higher it is, the better the reconstructed signal is. The objective speech quality measure (QC) would ideally reach to 1.000 with a perfect reconstruction.
\subsection{Software environment} \label{sec:SoftE}
The RF and other algorithms are executed using MATLAB (R2022a); for the LSTM implementation, we used Python (Keras, Pandas, Scipy, Numpy and TensorFlow libraries). The LSTM properties are as follows:
model (Sequential()), Dropout (0.2), Optimizer (Adam), learning rate (0.0001), and the loss
function for our model was measured using MSE (mean squared error). Finally, we fit the models with a batch size of 32 and 40 epochs (to avoid overfitting, as the training process had a stable loss within the first five iterations).
\section{Results and Analysis}
\label{sec:Results}
\begin{figure*}
\centering
\includegraphics[width=\textwidth]{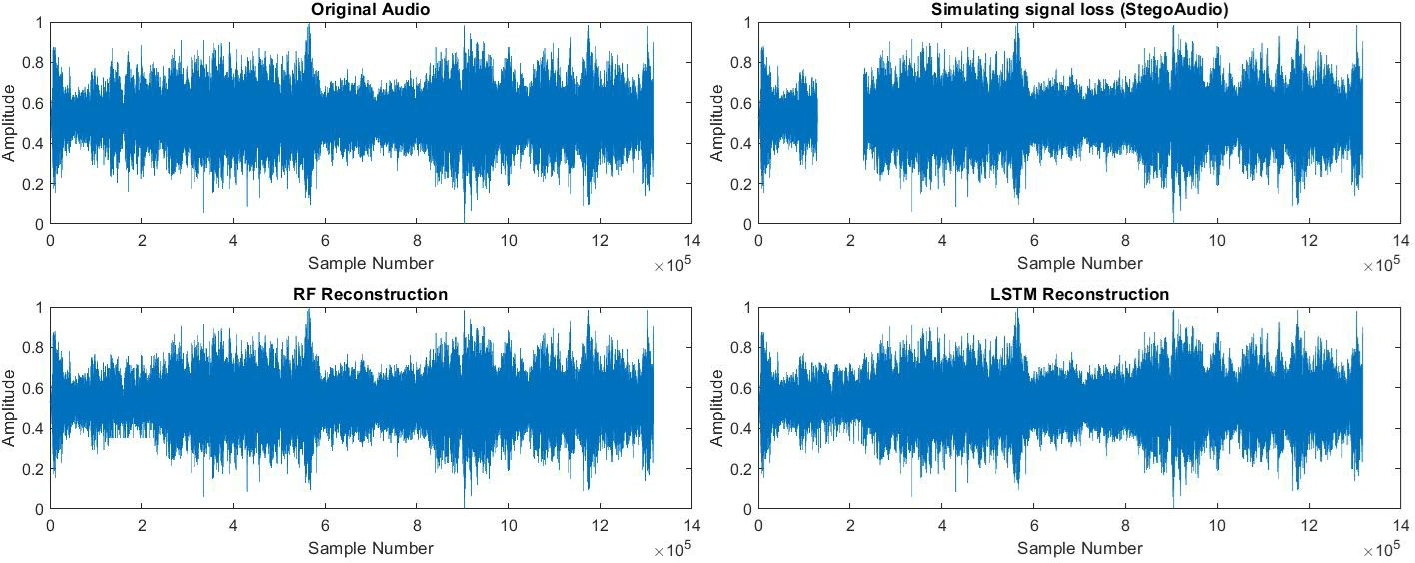}
\caption{Reconstruction of a short audio signal using RF and LSTM.} \label{fig4}
\end{figure*}

\begin{figure*}
\centering
\includegraphics[width=\textwidth]{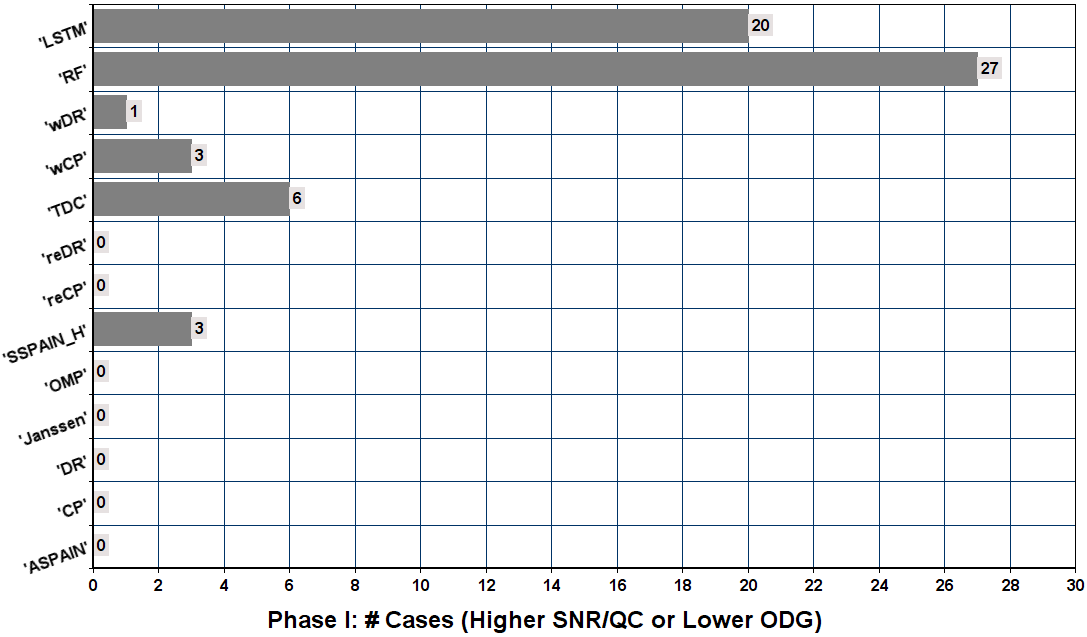}
\caption{Phase I: Determining the best performing non-learning methods. From the results of 20 tests on each of these methods, we have two competitive methods (CP and TDC) that will be tested in Phase II. Although ASPAIN has had no winning cases, since it is a recent algorithm, we opt to upvote it for relevance to phase II (see subsection ~\ref{sec:Dataset}). The \textit{X-axis} The X-axis denotes the number of cases a given algorithm outperforms other algorithms in terms of SNR or ODG (20 tests were measured). Detailed numerical results, on which this figure is based, are furnished in the supplementary files (\url{https://github.com/ARDISDataset/ARLAS/tree/main/Excel_Sheet}).} \label{fig5}
\end{figure*}

\begin{figure*}
\centering
\includegraphics[width=\textwidth]{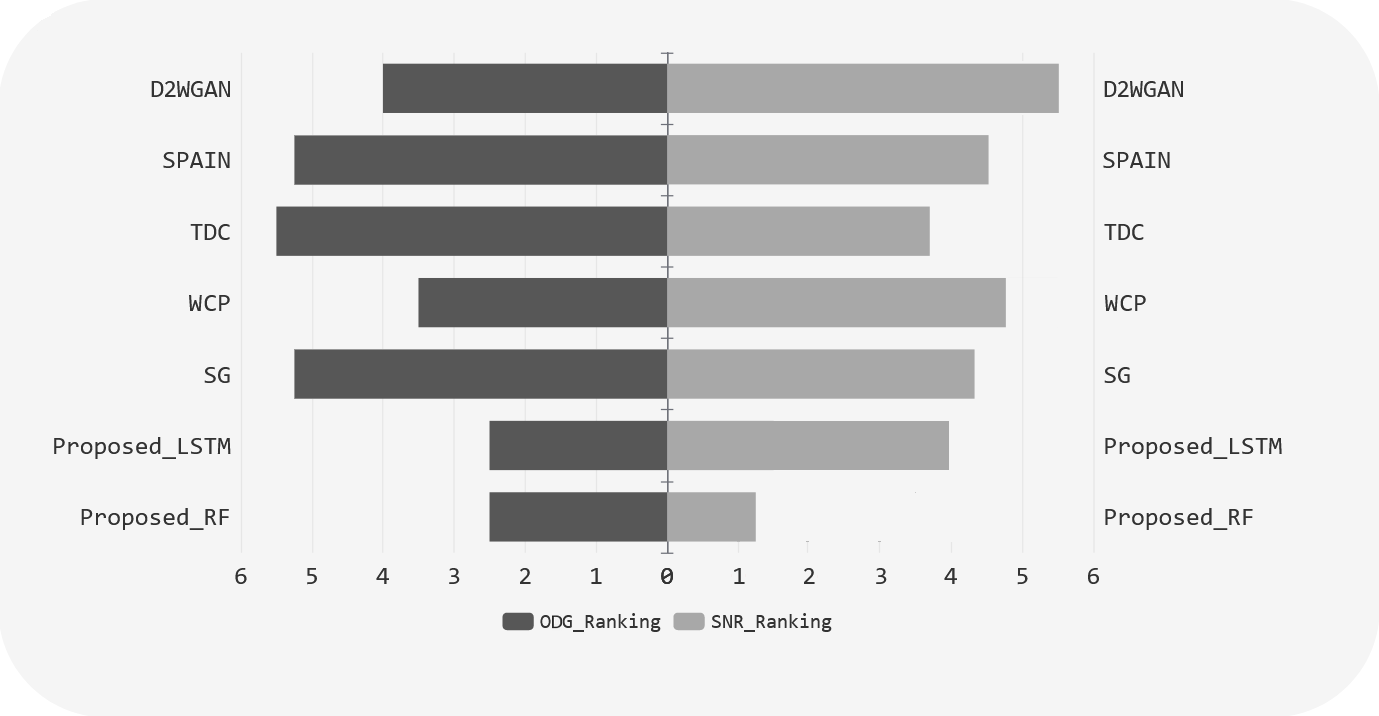}
\caption{Phase II: Determining the best-performing methods, including deep learning-based (D2WGAN) and graph-based (SG). From the results of 8 tests on each of these methods, we can observe that RF and LSTM (sequence-to-sequence modelling) exhibit promising results on both metrics (SNR and ODG), which ranked them at the top of the list. Detailed numerical results, on which this figure is based, are furnished in the supplementary files (\url{https://github.com/ARDISDataset/ARLAS/tree/main/Excel_Sheet}).} \label{fig6}
\end{figure*}

\begin{figure*}
\centering
\includegraphics[width=130mm]{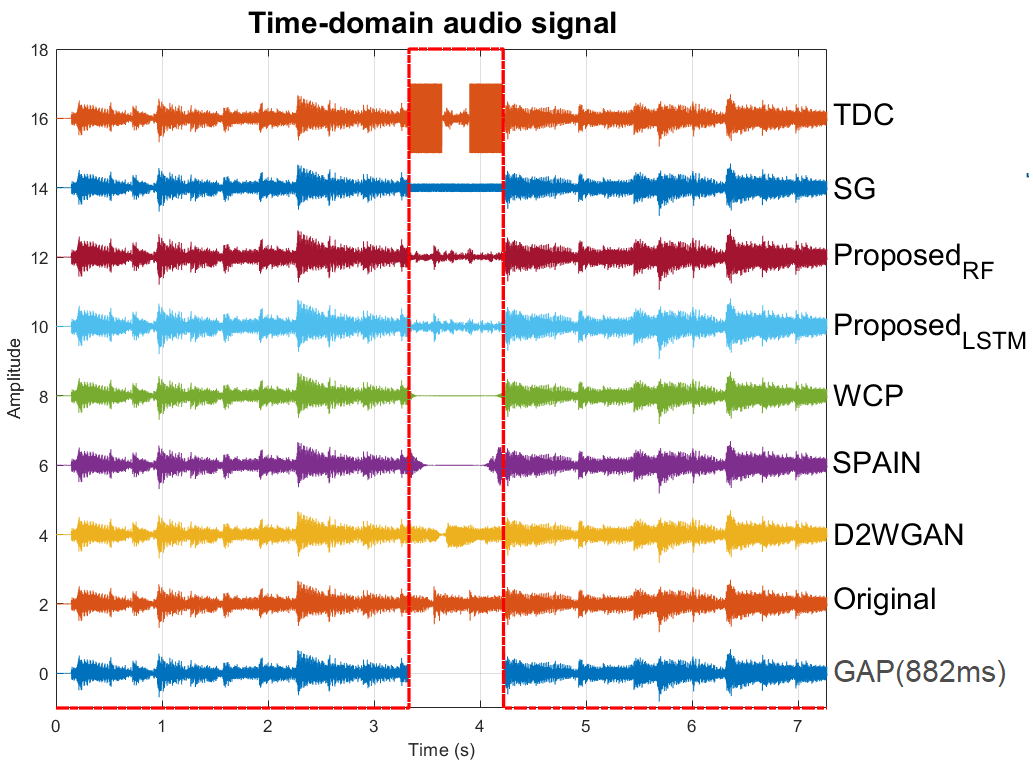}
\caption{Phase II - Audio reconstruction test-: Performance of the different audio inpainting algorithms on an example audio sample (\textit{extend-solo--1-real-36}) ~\cite{Ebner20}. The GAP signal is merely the Original audio sample but with the gap (highlighted in red dashed line) simulating dropped signal. } \label{fig7}
\end{figure*}

\begin{figure*}
\centering
\includegraphics[width=130mm]{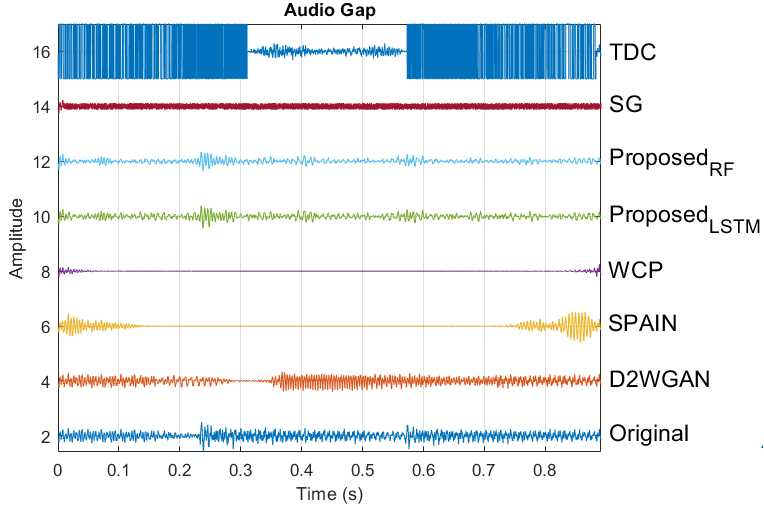}
\caption{A zoom-in into the reconstructed gap of Fig. ~\ref{fig7}. Notice how RF and LSTM predict the level of the amplitude despite the length of the gap (i.e., 882ms). } \label{fig8}
\end{figure*}

The comparison is performed between the original and the reconstructed audio signals in this study. After extracting the required training data from the sequence (see Fig.~\ref{fig4}), the data is then augmented using a scalogram. The aggregated data (features) are then passed to RF and LSTM models for training; see Algorithm ~\ref{alg:Two}. The test set is the hidden data embedded in the stego-audio whose dynamic range is extended using 2D Gaussian filter and whose feature space is extended using the scalogram; see Algorithms ~\ref{alg:Two} and ~\ref{alg:Three}. The evaluation of the reconstructed signals to the original signal (which acts as the reference for validation) is observed by calculating the statistical metrics reported in Sec. \ref{Metrics}. Many state-of-the-art audio inpainting algorithms (in Phase I) had their performance deteriorating when exceeding the range of very short gaps ($\approx$ 45ms); see Fig. ~\ref{fig5} where we show a visual summary of the performance with gaps of 100ms and 300ms (20 tests for each method).
In Phase II, we observed good performance of the proposed approach for more extended gap reconstruction (e.g., 800ms). Fig.~\ref{fig6} depicts a ranking summary, and Fig.~\ref{fig7} shows a real example of an audio gap reconstruction whose blow-up is presented in Fig.~\ref{fig8}, allowing closer scrutiny of the reconstruction quality. Additional examples are furnished online \footnote{Audio Clips (medium gaps): \url{https://ardisdataset.github.io/ARLAS2/}}. Moreover, the ability to handle lengthy missing gaps (e.g., 4000ms, 8000ms) teases our approach apart from other methods. The already existing methods, which we examined, are not designed nor able to handle these large gaps' reconstruction. The audio files of our experiments (i.e., LSTM, RF) are available online \footnote{Audio Clips (lengthy gaps): \url{https://ardisdataset.github.io/ARLAS/}}. The experiments reinforce that deep learning reconstruction approaches can benefit from embedded side-information if designed carefully, compare D2WGAN to our approaches in Fig. ~\ref{fig6} and  Table. ~\ref{tab2}.
\begin{table}
\caption{Average ranking of the different methods in Phase II (8 tests) measured using the three audio quality metrics. ODG and SNR are depicted graphically in Fig.~\ref{fig6}.}\label{tab2}
\begin{center}
\begin{tabular}{|l|l|l|l|}
\hline
Method & Rank-ODG  &  Rank-QC  & Rank-SNR\\
\hline
{\bfseries Proposed-LSTM} &  2.50 & 2.50 & 4.00\\
{\bfseries Proposed-RF} &	2.50	& 2.00	& 1.25 \\
{\bfseries wCP~\cite{Chambolle11}}	& 3.50 & 	5.25 &	4.75 \\
{\bfseries D2WGAN~\cite{Ebner20}}	& 4.00	& 2.75	& 5.50\\
{\bfseries SG~\cite{Perraudin18}}	& 5.25	& 5.75	& 4.25\\
{\bfseries SPAIN~\cite{Mokry19}} 	& 5.25	& 4.50	& 4.50\\
{\bfseries TDC~\cite{Mokry20}}	& 5.50	& 5.50	& 3.75\\
\hline
\end{tabular}
\end{center}
\end{table}
\section{Conclusions}
\label{sec:Conc}
The aim of this paper is to put forward a new framework which proposes the fusion of audio dither-based steganography with machine/deep learning for the active reconstruction of lost signals. The results show that the proposed solution is feasible and can enhance the reconstruction of lost audio signals (readers may wish to listen to the audio online, see URL in section ~\ref{sec:Results}). We conducted experiments on several types of signal drops of (100ms, 300ms), (500ms to 800ms) and (4000ms, 8000ms (shown online)). As a proof-of-concept, we can assert that, in general, the LSTM and the RF models are good models to utilise. Our approach is not meant to replace current inpainting audio methods but rather to assist them by providing latent side information. It can also benefit security systems in protecting audio files from unauthorised manipulation. The paper supplies extensive experiments, which we believe are compelling evidence of the efficacy of our proposed approach, a corollary when combining halftoning, steganography and machine learning.
To our knowledge, we found no similar implementation in the literature for audio missing-segment reconstruction. Thus, we conclude this paper by stating that the fusion of steganography and state-of-the-art machine learning algorithms can be considered for the active reconstruction of audio signals. However, as we pointed out in the discussion section, there is room for enhancement, for example, enhancing the algorithm for inverse-halftoning, which is an ill-posed problem.

%
%

\end{document}